\documentclass[useAMS,usenatbib]{mnras}
\usepackage{dblfloatfix}
\usepackage[titletoc]{appendix}
\usepackage{psfig, epsf, epsfig, mathtools,graphicx, pagecolor,amsmath}
\usepackage{float}
\usepackage{caption}
\usepackage{subcaption}
\usepackage{placeins}
\graphicspath{ {./images/} }

\title[Removing seeing effects from images]{
SeeingGAN: Galactic image deblurring with deep learning for better
morphological classification of galaxies} 
\author[F. K. Gan and others]
{Fang Kai Gan, Kenji Bekki and Abdolhosein Hashemizadeh\\ 
ICRAR M468
The University of Western Australia
35 Stirling Hwy, Crawley
Western Australia 6009, Australia}

\DeclareMathOperator{\SSIM}{SSIM}
\DeclareMathOperator{\MSE}{MSE}
\DeclareMathOperator{\PSNR}{PSNR}
\DeclareMathOperator{\CWSSIM}{CW-SSIM}
  
\begin{document}
\pagecolor{white}


\maketitle

\label{firstpage}

\begin{abstract}

Classification of galactic morphologies is a crucial task in galactic astronomy, and identifying fine structures of galaxies (e.g., spiral arms, bars, and clumps) is an essential ingredient in such a classification task. However, seeing effects can cause images we obtain to appear blurry, making it difficult for astronomers to derive galaxies' physical properties and, in particular, distant galaxies. Here, we present a method that converts blurred images obtained by the ground-based Subaru Telescope into quasi Hubble Space Telescope (HST) images via machine learning. Using an existing deep learning method called generative adversarial networks (GANs), we can eliminate seeing effects, effectively resulting in an image similar to an image taken by the HST. Using multiple Subaru telescope image and HST telescope image pairs, we demonstrate that our model can augment fine structures present in the blurred images in aid for better and more precise galactic classification. Using our first of its kind machine learning-based deblurring technique on space images, we can obtain up to 18\% improvement in terms of CW-SSIM (Complex Wavelet Structural Similarity Index) score when comparing the Subaru-HST pair versus SeeingGAN-HST pair. With this model, we can generate HST-like images from relatively less capable telescopes, making space exploration more accessible to the broader astronomy community. Furthermore, this model can be used not only in professional morphological classification studies of galaxies but in all citizen science for galaxy classifications.
\end{abstract}


\section{Introduction}

Classification of galactic morphologies has long been a critical task in extragalactic astronomy, not only because global galactic morphologies such as bulge-to-disk-ratios and spiral arm shapes can have fossil information of the galaxy formation, but also because the detailed statistical studies of galactic properties for each category can provide insights into the formation processes of different types of galaxies.  Galaxy classification schemes proposed in previous  pioneering works (e.g., \citealp{Hubble1936,Sandage1961,Vaucouleurs1991}) have long been used as standard tools in many observational and theoretical studies of galaxy formation and evolution. These days, galaxy classification is also done by non-professional astronomers such as the Galaxy Zoo project (e.g.,\citealp{Lintott2008,Bamford2009}), in which a large number of galaxy images ($>10^6$) from SDSS are provided for citizen science.

Galaxy classification has always been done by the human eye and will be done in future works. More recently, however, this process has begun to be automated by applying machine learning algorithms to actual observational data. For example, convolutional neural networks (CNNs) have been used in the automated classification of galactic morphologies for many galaxies (e.g., \citealp{Dieleman2015,Huertas-Company2015,Dom2018}). Galaxy classification using these deep learning algorithms has been successfully done for a large number ($>10^6$) of images from large ground-based telescopes such as the Subaru 8m telescope (e.g., \citealp{Tadaki2020}). Such quick automated classification is now considered to be the primary (and possibly only) way to classify a vast number of galaxies from ongoing and future surveys of galaxies such as LSTT and EUCLID.

One of the potential problems in classifying galaxy images from ground-based telescopes is that the images can be severely blurred owing to the seeing effects of the sky. Fine structures of galaxies, such as bars, spiral arms, and rings, is used to classify and quantify galaxies (\citealp{Nair2010}), such structures can be much less visible in galaxy images from ground-based telescopes, in particular, for distant galaxies (Fig. 22 in the paper demonstrates that the detection rates of bars and inner and outer rings depend strongly on seeing). Thus, If this optical blurring due to sky seeing can be removed by applying machine learning algorithms to real galaxy images,  it will provide significant benefits both to professional astronomers and non-professional ones who are working on the Galaxy Zoo project. 

The algorithm that we propose is based on Generative Adversarial Networks (GANs), which was originally proposed by \citealt{Goodfellow2014}. This technique is widely used in different image-related tasks such as style transfer \citep{Li2016} and super-resolution \citep{Ledig2016}. This solution has also been experimented with in the context of space astronomy by \citealt{Schawinski2017}, where they use a GAN-based network to remove noise from degraded galactic images, and more recently, galaxy image reconstruction by \citealt{Buncher2021}. Thus it is promising for us to develop a similar GAN-based model for deblurring galaxy images from ground-based telescopes.

The purpose of this paper is thus to develop a new GAN-based model that can convert blurred ground-based images of galaxies into clear HST-like galaxy images. Galaxy images from the HST do not have such problems as seeing effects because atmospheric distortion due to light travelling through the turbulent atmosphere is not a problem in these observations by a space telescope. In the present study, we use a large number of image pairs from the Subaru telescope (influenced by seeing) and from HST (not influenced by seeing at all) in order to generate the training data sets for our new GAN-based model (referred to as ``SeeingGAN'' from now on for convenience). We apply SeeingGAN to unknown data sets in order to quantify its prediction accuracy. Although the new algorithms presented here can be applied to galaxy images from the Subaru, other similar algorithms can be developed for the conversion of galaxy images from other ground-based telescopes (e.g., VLT etc.).

Space-image deblurring is not a new problem, and advanced ground-based telescopes utilise a technique called adaptive optics, where the mirrors in the telescopes can correct distortion in real-time by altering the shape of mirrors. It compares the way light is distorted when taken by a reference guiding star. In the present paper, we propose a totally different solution that avoids adaptive optics on large ground-based telescopes.

The plan of the paper is as follows. We describe our new GAN-based model's architectures for deblurring galaxy images in section \ref{section:network-characteristics}. We present the new GAN model results and quantify the prediction accuracies of the model in section \ref{section:results}. We summarise the conclusions of the results in section \ref{section:conclusion}.

\section{Network Characteristics}
\label{section:network-characteristics}

GAN networks were first designed by Ian Goodfellow \citep{Goodfellow2014}. A GAN network consists of two parts, a discriminator D and a generator G that forms a mini-max game. The generator learns to produce an artificial image to fool the discriminator, while the discriminator learns to distinguish between authentic images and artificial images synthesised by the generator. This network encourages the generator to produce a highly realistic solution to try to fool the discriminator.

\begin{equation}
\begin{multlined}
\min_{G} \max_{D} V(D,G) = E_{x\sim P_{data}(x)}[log D(x)] + \\
E_{z\sim P_{z}(z)}[log (1- D(G(z)))]
\end{multlined}
\end{equation}

Where $p_{data}$ is the distribution of the data, $p_{z}$ is the input noise generation distribution, $D(x)$ is the probability of data coming from real data more than generator. The GAN network attempts to maximise $D(x)$ an minimise $log(1-D(G(z)))$.

\subsection{Wasserstein GAN} 
\label{section:wgan}

In order to avoid problems such as mode collapse, unconvergence, etc. as noted in \citep{Salimans2016} with vanilla GAN models, we utilise the Wasserstein GAN \citep{Arjovsky2017} variant instead. Instead of using a discriminator to output the probability of a real/fake output, WGAN aims to score the image based on the “realness” and “fakeness” of an image. As compared to a vanilla GAN that utilises Jensen-Shannon Divergence to measure the real/fake distributions, WGAN seeks to measure the differences using 1-Wasserstein distance (Earth-Mover distance) \citep{Arjovsky2017}. Intuitively, Wasserstein loss provides a proportional metric to relate the predicted and expected probability distribution performance that can be back-propagated for training.

\subsection{Conditional GAN}
In conventional GAN networks, the job of the discriminator is to classify the authenticity of images. However, given that if the dataset has additional auxiliary information, these data can be fed into the generator network in parallel to improve the GAN network. As summarised by \citealp{Isola2016}, conditional GANs (cGAN) can be used to train general networks to learn the mapping from the input image and the random noise vector to the output. Intuitively, these networks are conditioned by the additional input data to produce an output in the desired class \citep{Mirza2014}. In contrast to regular GAN networks that map a random noise vector $z$ to the output image $y$, $G:\{z\} \rightarrow y$, conditional GANs learn the mapping from the given image {x} and a noise vector {z} to the output $y$, $G:\{x, z\} \rightarrow y$. Where $x$ can be any auxiliary information, such as class labels. In our case, it is the observed low-resolution image.

\subsection{Skip Connections}
\begin{figure*}
\includegraphics[width=\textwidth]{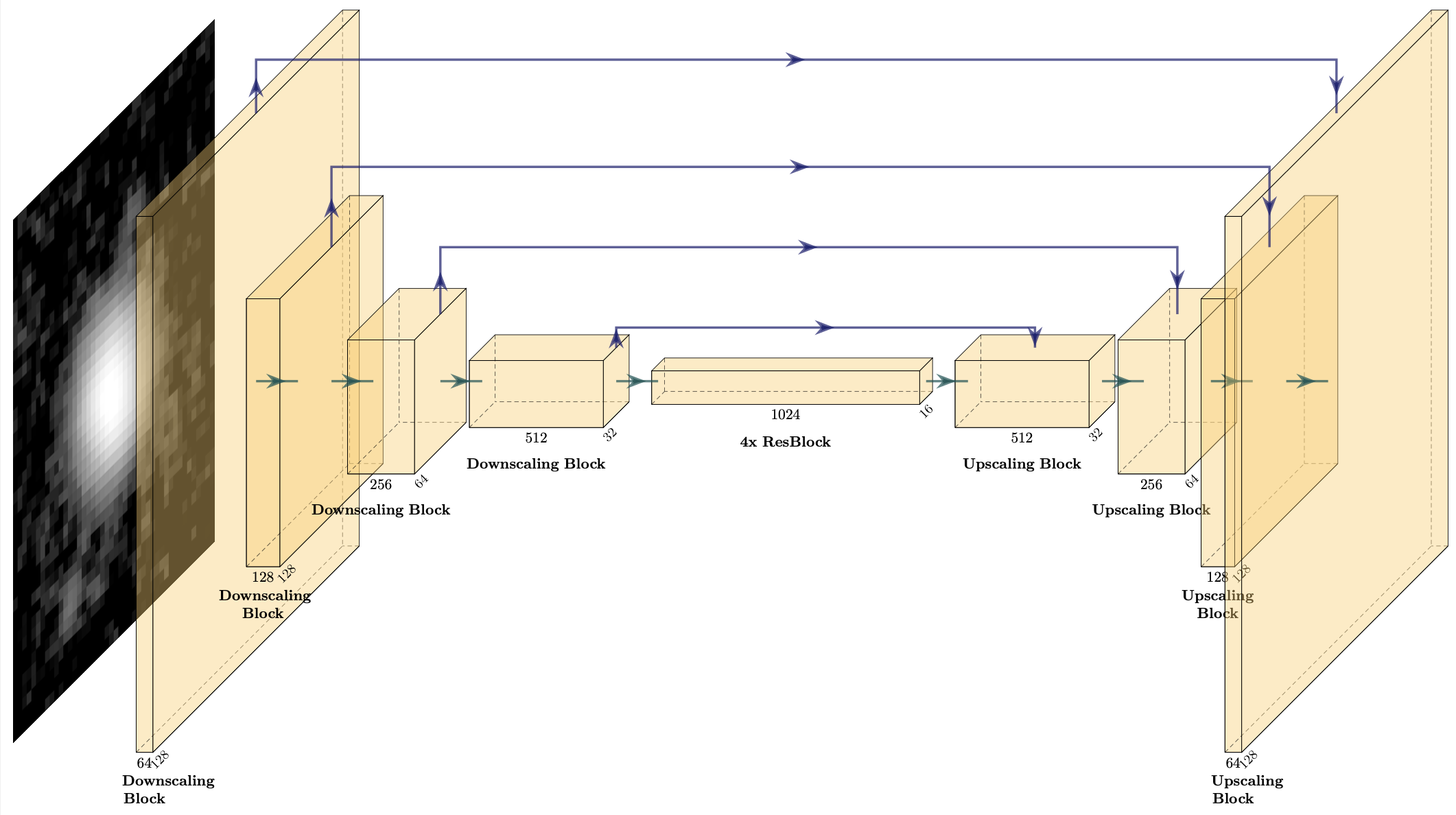}
\caption{Generator Architecture}
\label{section:generator}
\end{figure*}

\begin{figure}
\includegraphics[width=\textwidth/2-50pt]{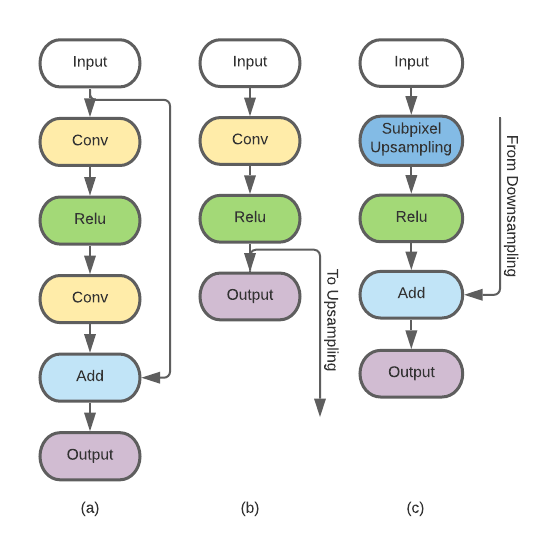}
\caption{(a) Modified ResNet block; (b) Downscaling block; (c) Upscaling block}
\label{section:generator-component}
\end{figure}

We used skip connections in our model to make it a U-Net like encoder-decoder model similar to the \textit{Pix2Pix} model \citep{Isola2016}. Generally, a more significant number of CNN layers will increase model performance. However, this method's drawback is that deeper networks are often difficult to train due to the vanishing gradient effect. As the model tries to estimate a gradient to update the weights during back-propagation, due to the depth of the model, the gradients become so small where it has almost no effect when it reaches the input layer. To combat this effect, several solutions such as multi-level hierarchy networks proposed by Jurgen Schmidhuber \citep{Schmidhuber1992}, where each layer is trained layer by layer; or using other activation functions such as ReLU \citep{Glorot2011}, where they are less susceptible to the vanishing gradient problem.

Our paper will utilise one of the newer solutions pioneered by Kaiming He et al. in 2015 – ResNets \citep{He2015}. ResNets are regular networks with skip connections. Skip connections or “shortcut connections” connects a lower layer to a higher layer bypassing intermediate layers. This proves to be very powerful because these skip-connections acts as a “highway” for deeper layers to quickly learn by reducing the impact of vanishing gradients and then gradually learning the skipped layers to to optimise the model further as the “highway” connections converge. Additionally, these skip connections allow additional features from the lower layer to flow into the higher layer.

\subsection{Subpixel Upsampling with ICNR}
\label{section:subpixel}

Convolution transpose is simply a deconvolution layer that performs the opposite of a convolution layer in the decoding network. However, the deconvolution layers commonly have checkerboard patterns as they can usually have uneven overlaps. In theory, weights can be learnt to avoid this problem, but networks often struggle to evade this issue.

We attempted to reduce this effect by using nearest neighbour upsampling and convolution layers, but the images we obtained tend to be blurred since the image is upsampled via an interpolation method.

Therefore, we used a technique proposed by Wenzhe Shi et al. \citep{Shi2016}  – Sub-Pixel Convolutional Neural Network. Instead of using zeros in-between pixels in regular deconvolution, we calculate more convolutions of the low-resolution image and resize the output map into the higher resolution layer. This technique avoids zero data in regular convolution transpose upsampling layers. This technique still suffers from checkerboard artefacts, albeit to a lesser extent.  To further eliminate checkerboard artefacts, we used a technique called ICNR (Initialised to convolution NN resize) proposed by \citealp{Aitken2017}. Instead of random weight initialisations, we set the weight matrix to the nearest-neighbour neural network resize before the sub-pixel upsampling layer. This completely eliminates the checkerboard effect.

\subsection{Perceptual Loss}

The loss function we utilised for this network is a combination of perceptual loss \citep{Johnson2016} and Wasserstein loss. The most widely used loss function for GAN networks is L2 loss (Mean Square Error, MSE). However, networks that utilise such losses often lead to blurry effects in the output image because the generator tends to output images to fulfil the pixel-wise average goal of MSE \citep{Isola2016}. 

Hence, we adopt a more robust solution called perceptual loss. It is still a MSE loss function, but it is a loss function of the CNN layer's feature maps. Our model used a pre-trained VGG-19 layer on ImageNet, using its’ $conv_{3,1}$  layer feature map. The abstraction of the VGG-19 layer can extract more important features from the image that are more representative perceptually, which makes the image more realistic. The selection of feature layers was an experimental procedure. We went through every feature layer and compared the results of using the said layers to select the most optimal feature layer. 

The other loss we utilised was the Wasserstein loss described in section \ref{section:wgan} above. This loss function provides a continuous distance metric between the predicted output and the original image distribution, which is used for back-propagation.

\subsection{Batch Normalisation}

We did not include the batch normalisation (BN) layer in the generator model commonly used in GAN networks. BN layers work by normalising a layer's input features to have zero mean and unit variance, which is known to stabilise the training process. However, due to the incorporation of the skip connections, the model can directly learn the feature mapping of the image pairs. Hence, normalising the feature becomes less crucial \citep{Shao2020}. Although BN layers are known to improve training time, the output tends to be suboptimal. As a result, we removed the BN layers to reduce memory consumption and improve model performance.

\begin{figure}
\begin{center}
\includegraphics[width=\textwidth/2-50pt]{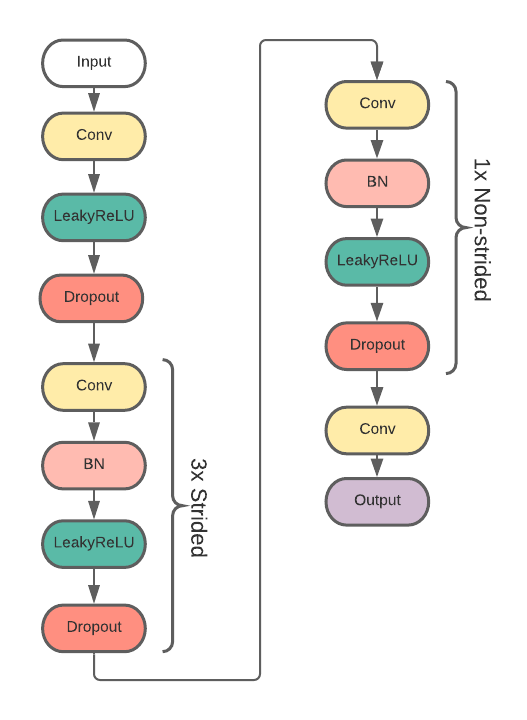}
\caption{Discriminator Architecture (PatchGAN) \citep{Isola2016}}
\label{section:discriminator}
\end{center}
\end{figure}

\subsection{Network Architecture}
The generator network consists of 4 strided convolution blocks followed by 4 modified ResNet blocks and 4 upsampling blocks to upscale the image back to the source size. The generator architecture is shown in Fig. \ref{section:generator}. On the strided downscaling convolution shown in Fig. \ref{section:generator-component} (b) , a skip connection is made to the opposite and symmetrical transposed convolution block, forming a U-Net architecture. The downscaling layers consist of a convolution block followed by a ReLU block. The modified ResNet blocks shown in Fig. \ref{section:generator-component} (a) consists of convolution block, followed by a ReLU block and another convolution block. Each block has a skip connection from the input to the output of each block. The Upsampling blocks shown in Fig. \ref{section:generator-component} (c) consist of Subpixel Upsampling layers as discussed in section \ref{section:subpixel}.

The discriminator network is the same as PatchGAN \citep{Isola2016}, shown in Fig. \ref{section:discriminator}. The network is formed with a convolution block followed by LeakyReLU and a dropout layer. Then 3 strided convolutions blocks are added (strided convolution, instance normalisation, LeakyReLU, dropout) and 1 non-strided convolution block is added (convolution, instance normalisation, LeakyReLU, dropout) and lastly a convolution block before flattening to a fully connected layer for the discriminator output.

\subsection{COSMOS ACS imaging data}
\label{sec:ACS/HST}

The Cosmic Evolution Survey (COSMOS) is arguably one of the most comprehensive deep-field surveys to date, covering a $\sim2$ deg$^2$ equatorial field of sky, designed to explore the evolution of the large scale structure (LSS) and the evolution of galaxies, AGN and dark matter out to $z\sim6$ \citep{Scoville2007}. The high-resolution HST I-band imaging (F814W) in COSMOS to a point source depth of $I_{AB} = 28$ mag taken by the Advanced Camera for Surveys (ACS) allows for unprecedented morphological and structural investigations. More than $\sim 2$ million galaxies are detected in the COSMOS region at the resolution of less than $100$ pc \citep{Scoville2007}. 

The COSMOS field is centred at RA = $10:00:28.600$ and DEC = $+02:12:21.00$ (J2000) and incorporates an extensive supplementary photometric and spectroscopic observations from various ground- and space-based telescopes providing a multi-wavelength data from X-ray to radio (see e.g., \citealt{Capak2007}; \citealt{Hasinger2007}; \citealt{Lilly2007}; \citealt{Sanders2007}; \citealt{Schinnerer2010}; \citealt{Davies2018}).   
In this study, we make use of the 1.7 deg$^2$ imaging in this field with the Advanced Camera for Surveys (ACS\footnote{ACS Hand Book: \href{http://www.stsci.edu/hst/acs/documents/handbooks/current/c05\_imaging7.html\#357803}{www.stsci.edu/hst/acs/documents/}}) on HST that was observed during 590 orbits in the F814W filter (I-band). This wavelength not only provides the almost complete coverage of the field but also samples a rest-frame wavelength suitable for the optical morphological studies of galaxies out to $z\sim1$ \citep{Koekemoer2007}. 
We use the drizzled ACS/F814W imaging that is resampled to the pixel scale of 0.03 arcsecond/pixel using the MultiDrizzle code (\citealt{Koekemoer2003}) while the raw ACS pixel scale is 0.05 arcsecond/pixel.
The frames were downloaded from the public NASA/IPAC Infrared Science Archive (IRSA) webpage\footnote{\href{https://irsa.ipac.caltech.edu/data/COSMOS/images/acs\_2.0/I/}{https://irsa.ipac.caltech.edu/data/COSMOS/images/acs\_2.0/I/} } as fits images.

In addition to the HST imaging data, in order to build our training data set (pairs of high- and low-resolution images of each galaxy) we use the companion ground-based imaging of our galaxies in $r^\prime$ band ($6213.0$ \AA) taken by the Suprime-Cam on the 8.2m Subaru telescope in the COSMOS field (\citealt{Taniguchi2007}). The Subaru Suprime-Cam imaging has a pixel scale of 0.202 arcsecond/pixel, i.e., $\sim6.7 \times$ HST/ACS pixel scale. 
We downloaded the Subaru imaging frames from publicly available COSMOS data base\footnote{\href{https://irsa.ipac.caltech.edu/data/COSMOS/images/subaru/}{https://irsa.ipac.caltech.edu/data/COSMOS/images/subaru/}}.

\subsection{Sample Selection and Postage Stamp Cutouts:} 
\label{subsec:SmplSel}

Despite HST's high spatial resolution in resolving galaxy substructures at very high redshifts ($z > 1$), it is still challenging as galaxies become too dim or small in angular size. Therefore to select a suitable subsample of galaxies to resolve their structures and cover all morphological types, we need to limit our sample to certain redshift and stellar mass ranges.
We select 700 galaxies out of the D10/ACS sample generated by Hashemizadeh et al. (in prep.) built upon the source detection and photometric measurements using {\sc ProFound} code \citep{Robotham2018} on the UltraVISTA Y-band imaging of the COSMOS field \citep{McCracken2012} as part of the input catalogue for Deep Extragalactic VIsible Legacy Survey (DEVILS, \citealt{Davies2018}). Hashemizadeh et al. provide a visual morphological classification of $\sim 36,000$ galaxies in the COSMOS field separating galaxies into bulge+disk, pure-disk, elliptical and irregular/merger categories. Using their morphological classification, we assure that our final training sample consists of all significant morphological types reducing the sample's. 
We then limit our redshift range to $z < 0.6$ and stellar mass to $M_{*} > 10^9 M_\odot$. Note that redshifts and stellar masses are extracted from COSMOS2015 catalog (\citealt{Laigle2016}). 

\begin{figure}
\includegraphics[width=\textwidth/2-20pt]{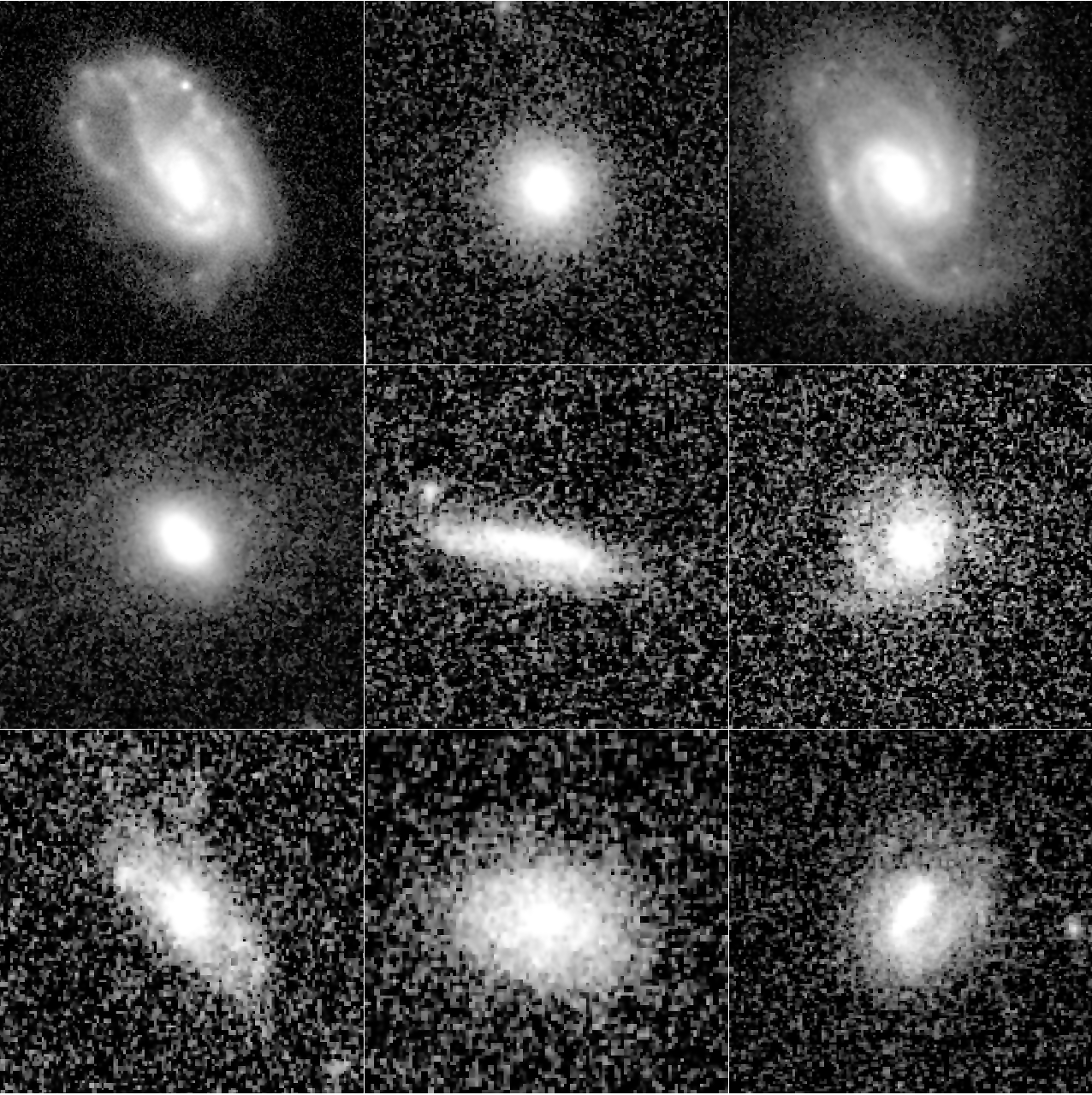}
\caption{9 sample HST images downloaded from NASA/IPAC Infrared Science Archive (IRSA) webpage, as described in section \ref{sec:ACS/HST}. These images were selected from the training dataset, labelled as the ground truth, where the model attempts to convert the images from Fig. \ref{section:subaru-image} to this.}
\label{section:hst-image}
\end{figure}

\begin{figure}
\includegraphics[width=\textwidth/2-20pt]{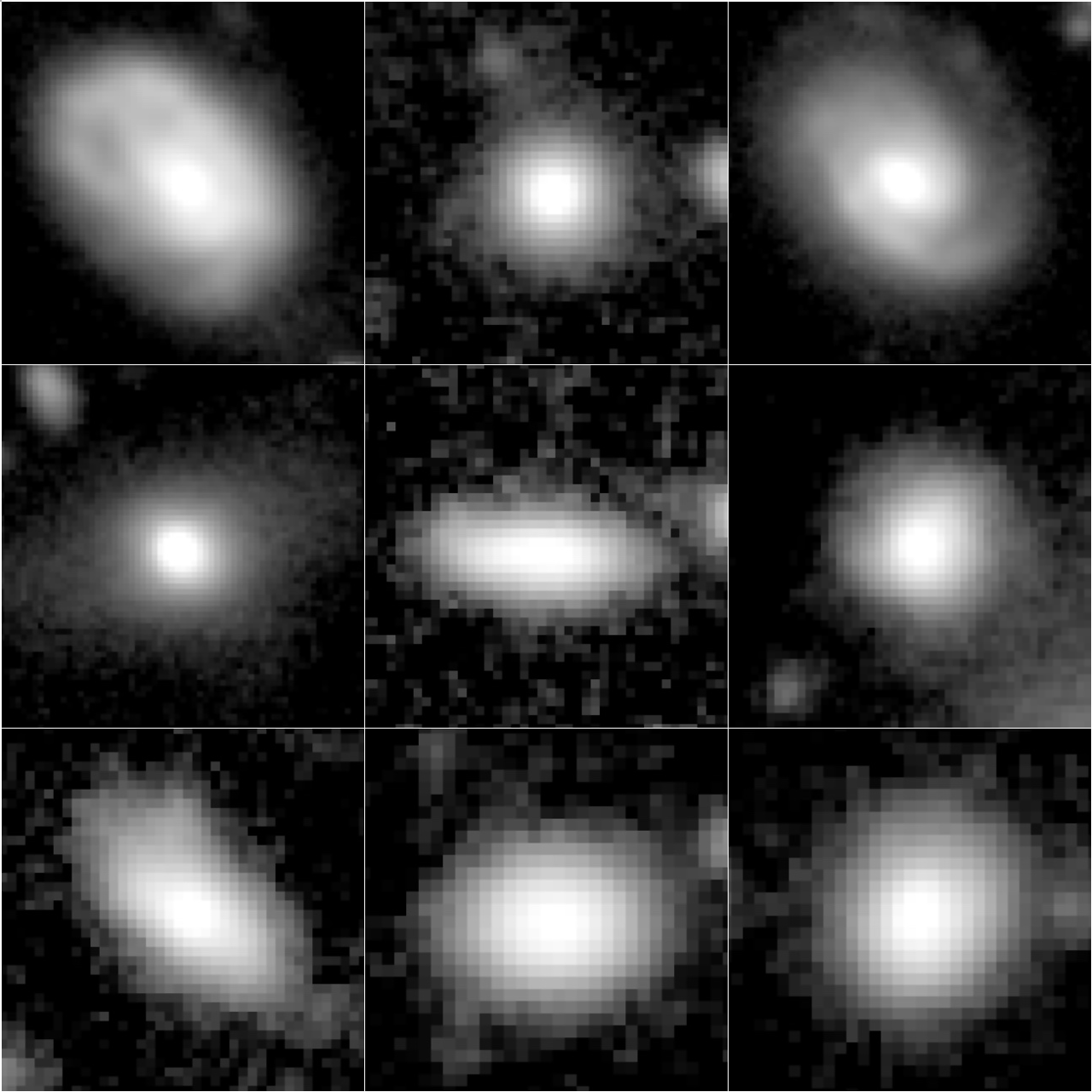}
\caption{9 sample images from the Subaru telescope, downloaded from the COSMOS project, as described in section \ref{sec:ACS/HST}. These images are the counterpart images of the HST images in Fig \ref{section:hst-image}. These images are selected from the training dataset and labelled as the source dataset, where the model attempts to convert these images to images in Fig \ref{section:hst-image}.}
\label{section:subaru-image}
\end{figure}

Finally, we generate postage stamps of these 700 galaxies in both imaging data (i.e., HST/ACS and Subaru Suprime-Cam). Figure \ref{section:hst-image} and \ref{section:subaru-image} show the postage stamps of nine of our galaxies in the HST/ACS and Subaru images, respectively. Instead of a fixed cutout size, the stamps' dimensions are elected to be $2 \times\mathrm{R90}$ on each side, where R90 is the elliptical semi-major axis containing 90\% of the total flux in UltraVISTA Y-band. With this dataset, the HST images were considered as ground truth while the images from the Subaru ground telescope were considered the source dataset.

\subsection{Evaluation Metrics}
\label{section:eval_metrics}

To quantify our model, we employed multiple metrics to measure model performance numerically. One major characteristic we would like to highlight is, even though image pairs from our dataset were spatially matched as closely as possible, the images may still not be perfectly aligned. Hence, attempts to measure pixel-to-pixel improvements was difficult due to the spatial variability of the images.

\begin{figure*}
 \begin{minipage}[l]{1.8\columnwidth}
        \centering
        \includegraphics[width=\textwidth]{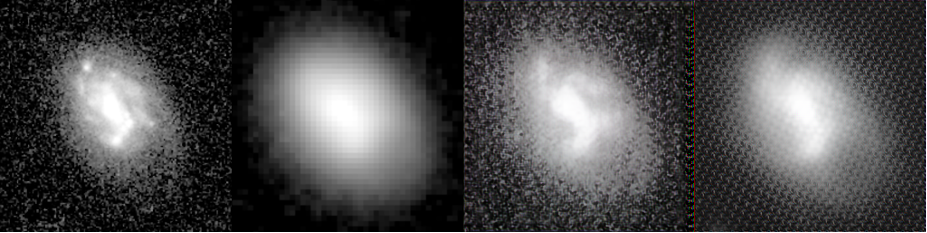}
\caption{Image pairs with different loss layers, from left - HST image, Subaru Image, SeeingGAN with VGG-19 $conv_{5,4}$ as loss layer,  SeeingGAN with VGG-19 $conv_{3,3}$ as loss layer. Different layers in the discriminator's VGG-19 extracts different features from the generator output. Hence, we iterated different layers ad compare the results to determine the most suitable layer (hence, feature) for the discriminator's loss layer.}
\label{section:loss-layers}
\end{minipage}
\end{figure*}
     
\subsubsection*{Peak Signal to Noise Ratio (PSNR)}

\begin{equation}
\begin{multlined}
\PSNR = 20 \log_{10} ( \frac{MAX_f}{\sqrt{MSE}})
\end{multlined}
\end{equation}

\begin{equation}
\begin{multlined}
\MSE = \frac{1}{mn}\sum_{0}^{m-1}\sum_{0}^{n-1}||f(i,j) - g(i,j)||^2
\end{multlined}
\end{equation}

The most common quality metric to compare the reconstruction quality of two images is PSNR. The lower the value of MSE, the smaller the pixel-wise differences, and thus the higher the PSNR value is. However, the drawback of PSNR is that it does not consider the qualitative measure of the image as it solely relies on the pixel-wise difference between 2 images. Thus, a higher PSNR does not necessarily equate to a better image per se, as it is not very well matched to perceived image quality \citep{Zhou2004}. 

\subsubsection*{Structural Similarity Index Metric (SSIM)}

As noted above, PSNR is not a very good metric to quantify image quality. Thus, SSIM was created by \citealt{Zhou2004}  to quantify image similarity better. The SSIM index is a perceptual metric that quantifies the image quality degradation proposed by Zhou et al. \citep{Zhou2004}. Given that two patches $x = \{x_i | i = 1, ... M\}$ and $y = \{y_i | i = 1, ... M\}$,

\begin{equation}
\begin{multlined}
\SSIM(x,y) = \frac{(2\mu_x\mu_y + C_1) + (2 \sigma _{xy} + C_2)}     {(\mu_x^2 + \mu_y^2+C_1) (\sigma_x^2 + \sigma_y^2+C_2)}
 \end{multlined}
\end{equation}

\noindent where $\mu$, $\sigma$ are the sample mean and standard deviation, while $C_1$ and $C_2$ are two constants to stabilise the weak denominator.

The index measures the perceptual difference between the two images. This algorithm considers the visible structures in the image by comparing the surrounding pixels of each section of the image. Due to this characteristic, it is considered a better indicator of image quality than PSNR due to its ability to quantify perceptual similarity. SSIM ranges between -1 and 1, where 1 is a perfectly similar image 0 indicates no structural similarity. However, the major drawback of this metric is, it is quite sensitive to scaling, translation and rotation \citep{Gao2011}.

\begin{figure*}
 \begin{minipage}[l]{1.8\columnwidth}
        \centering
        \includegraphics[width=\textwidth]{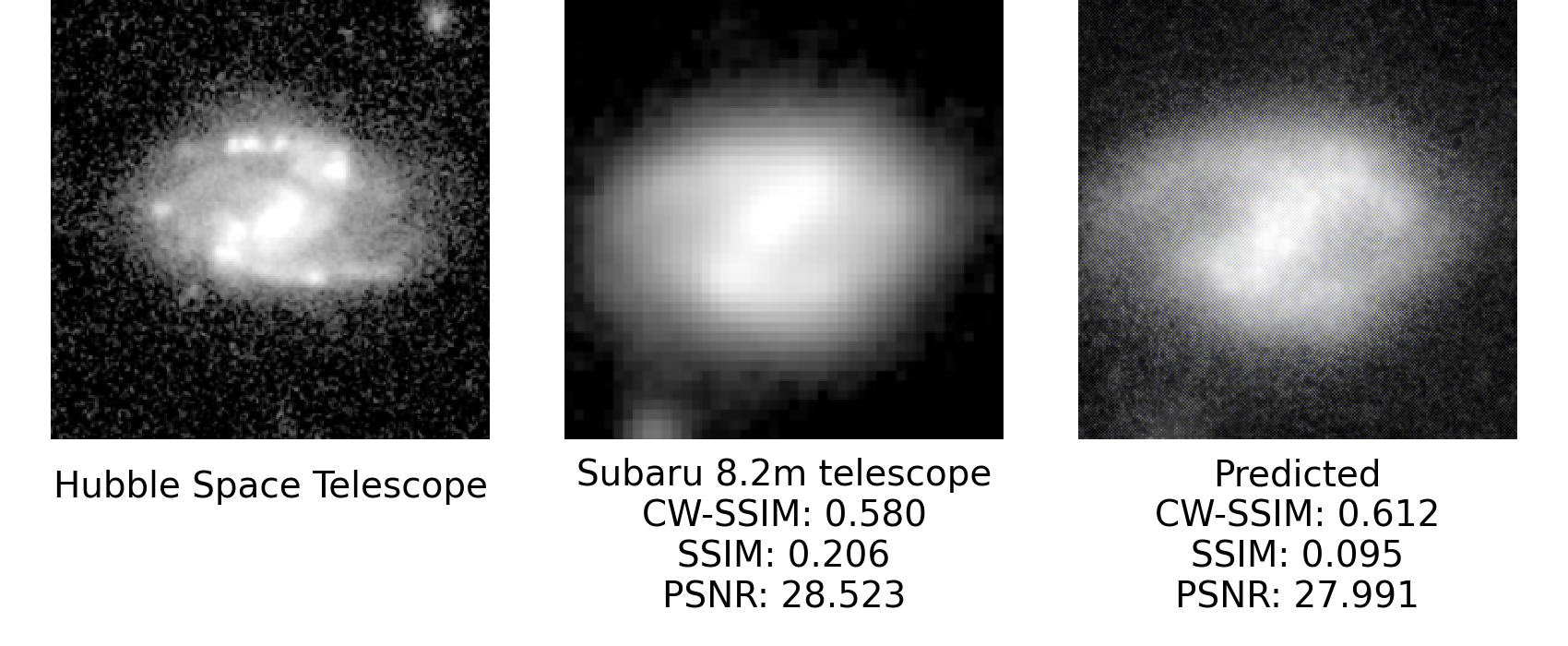}
 \caption{One enlarged sample result predicted by SeeingGAN. The predicted image is obtained by feeding the Subaru 8.2m telescope's image into SeeingGAN. The resultant image has a higher CW-SSIM score, which indicates a better similarity to the HST image.}
\label{section:result-enlarged}
\end{minipage}
\end{figure*}

\begin{figure*}
 \begin{minipage}[l]{1.0\columnwidth}
         \centering
        \includegraphics[width=\textwidth]{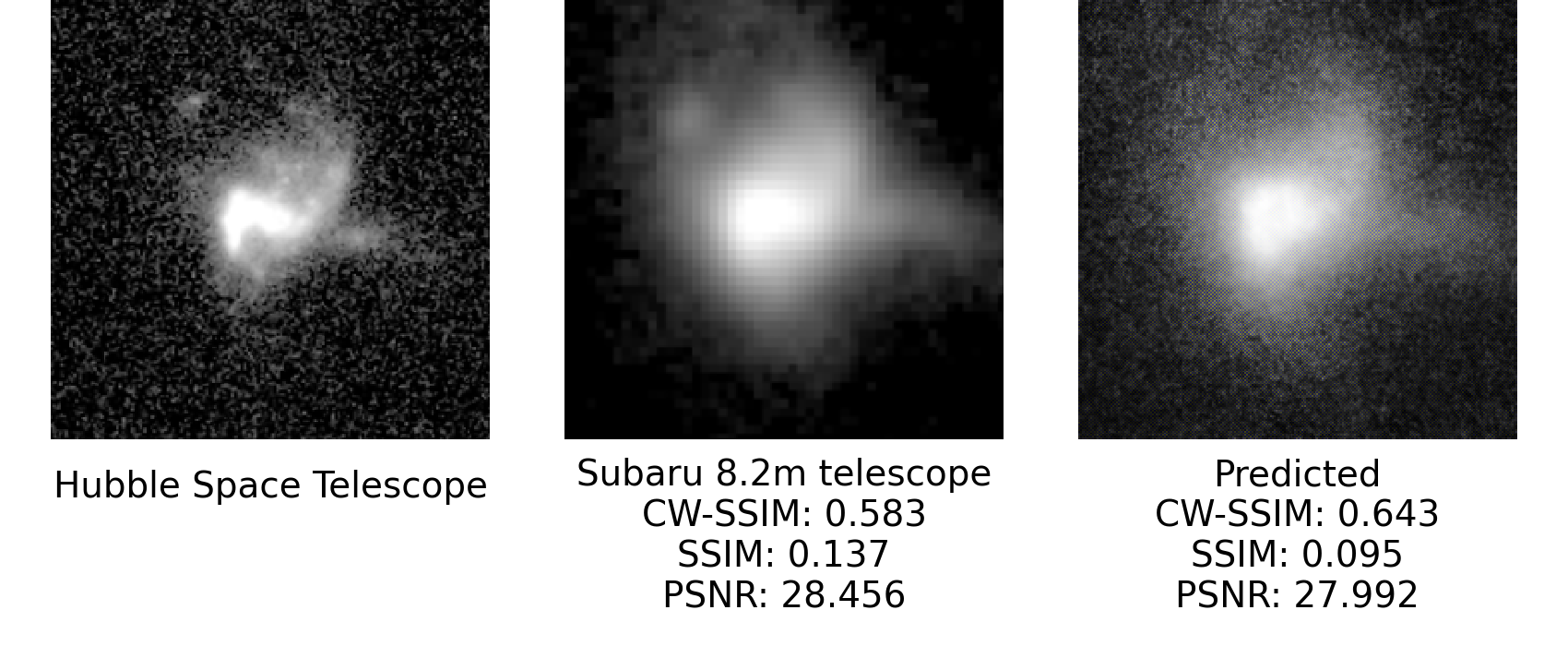}
         (a)
     \end{minipage} 
             \vspace{0.05in}
     \hfill{}
     \begin{minipage}[r]{1.0\columnwidth}
         \centering
	\includegraphics[width=\textwidth]{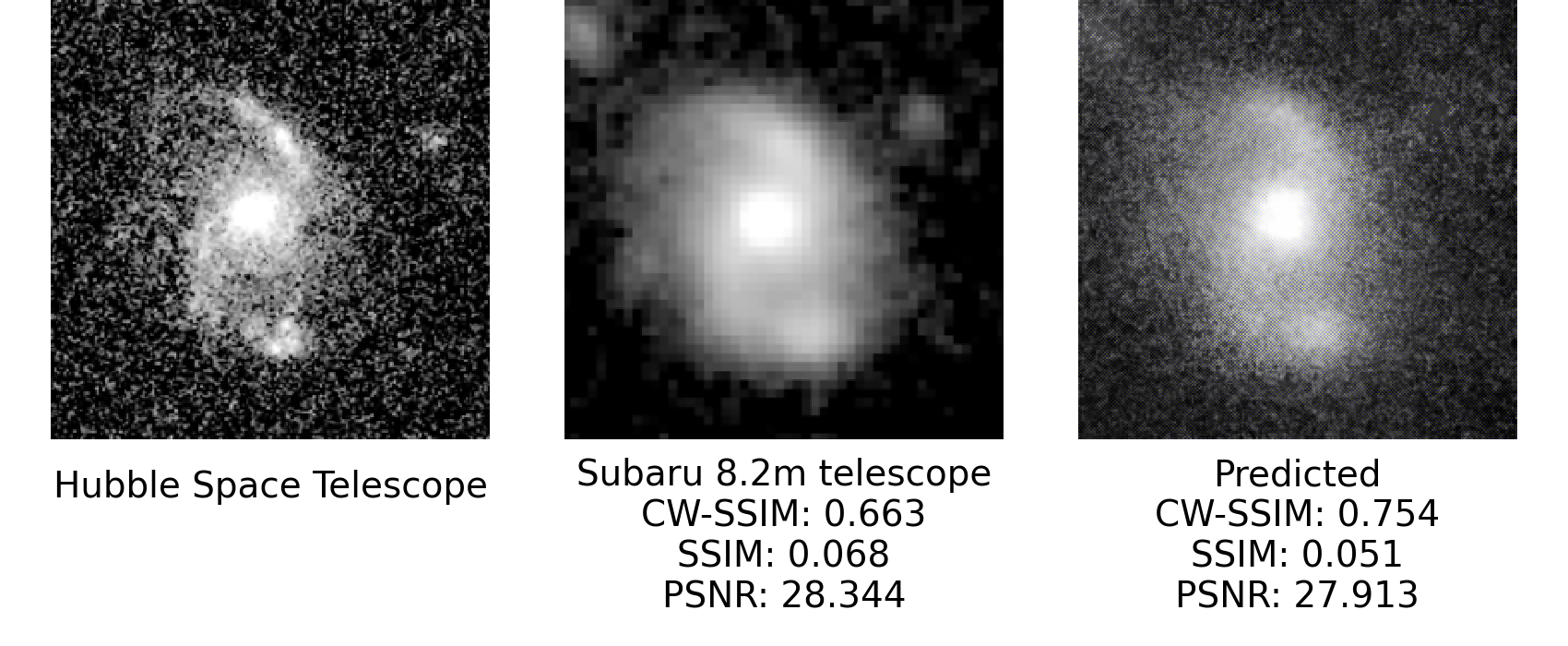}
	(b)
     \end{minipage}
        \vspace{0.05in}

      \begin{minipage}[l]{1.0\columnwidth}
         \centering
	\includegraphics[width=\textwidth]{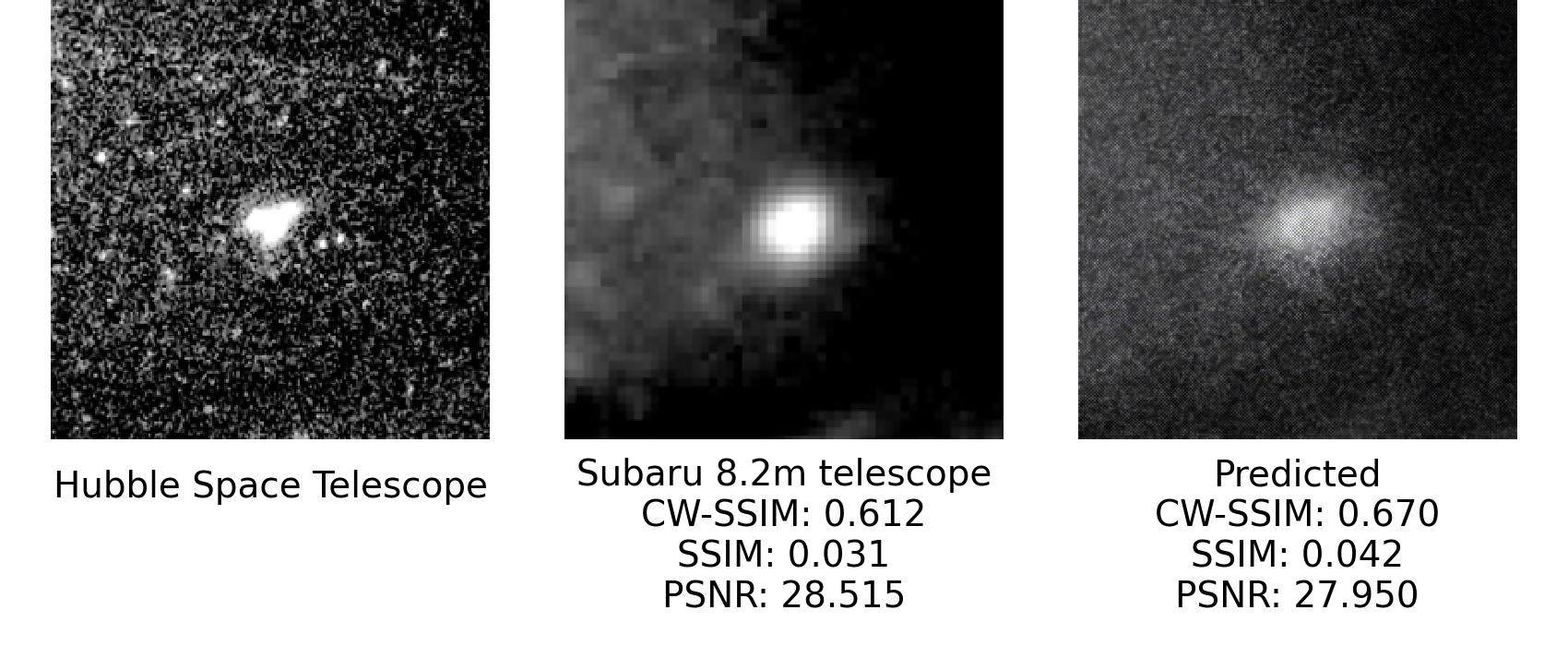}
	(c)
     \end{minipage}
             \vspace{0.025in}
          \hfill{}
      \begin{minipage}[r]{1.0\columnwidth}
         \centering
	\includegraphics[width=\textwidth]{s4}
	(d)
     \end{minipage}
     	        \vspace{0.05in}

	              \begin{minipage}[l]{1.0\columnwidth}
         \centering
	\includegraphics[width=\textwidth]{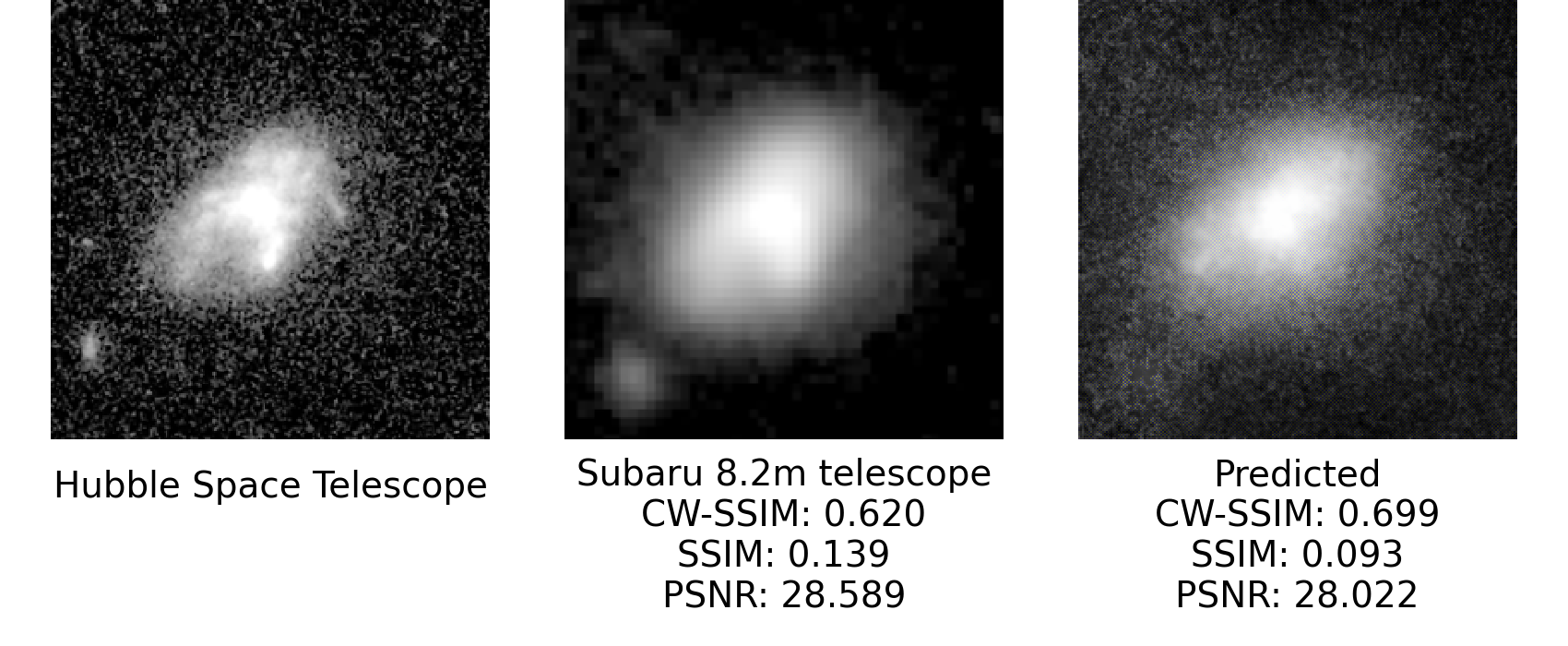}
	(e)
     \end{minipage}
             \vspace{0.025in}
          \hfill{}
      \begin{minipage}[r]{1.0\columnwidth}
         \centering
	\includegraphics[width=\textwidth]{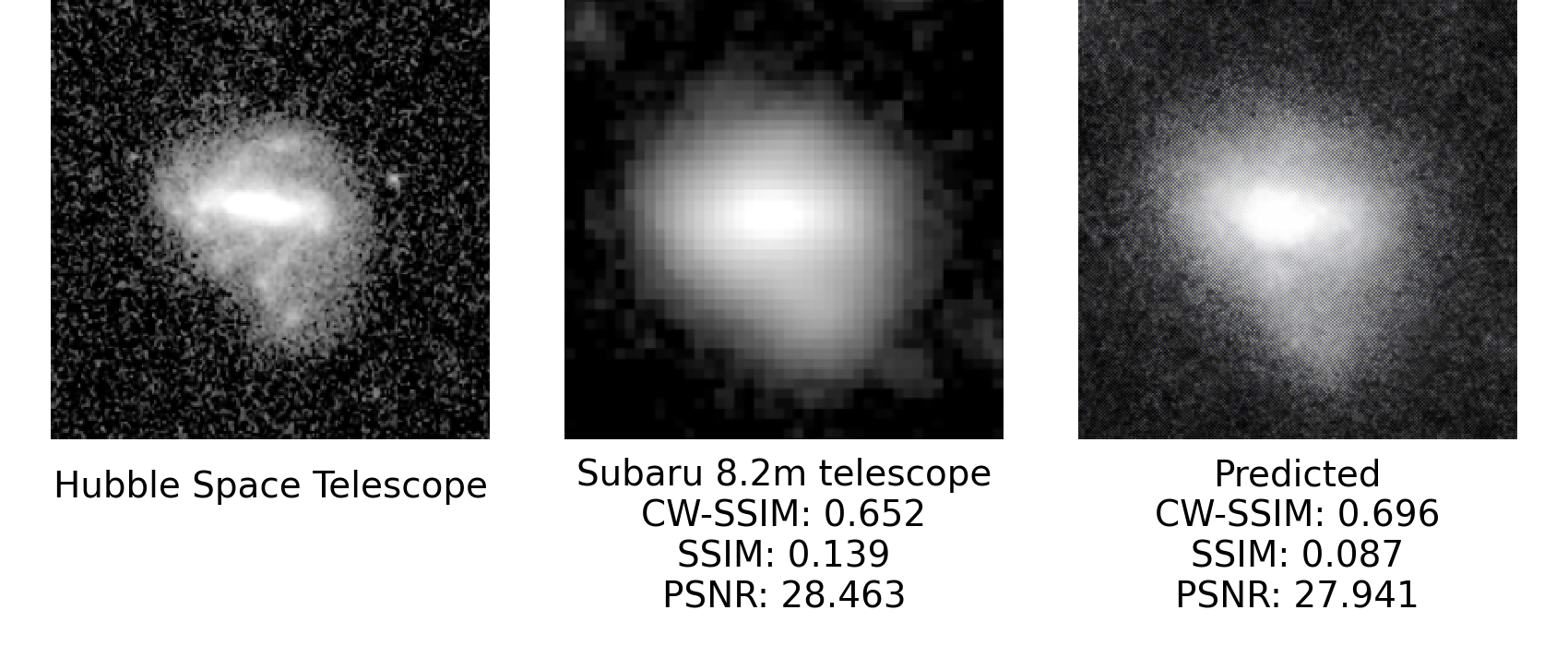}
	(f)
     \end{minipage}
     	        \vspace{0.05in}

	              \begin{minipage}[l]{1.0\columnwidth}
         \centering
	\includegraphics[width=\textwidth]{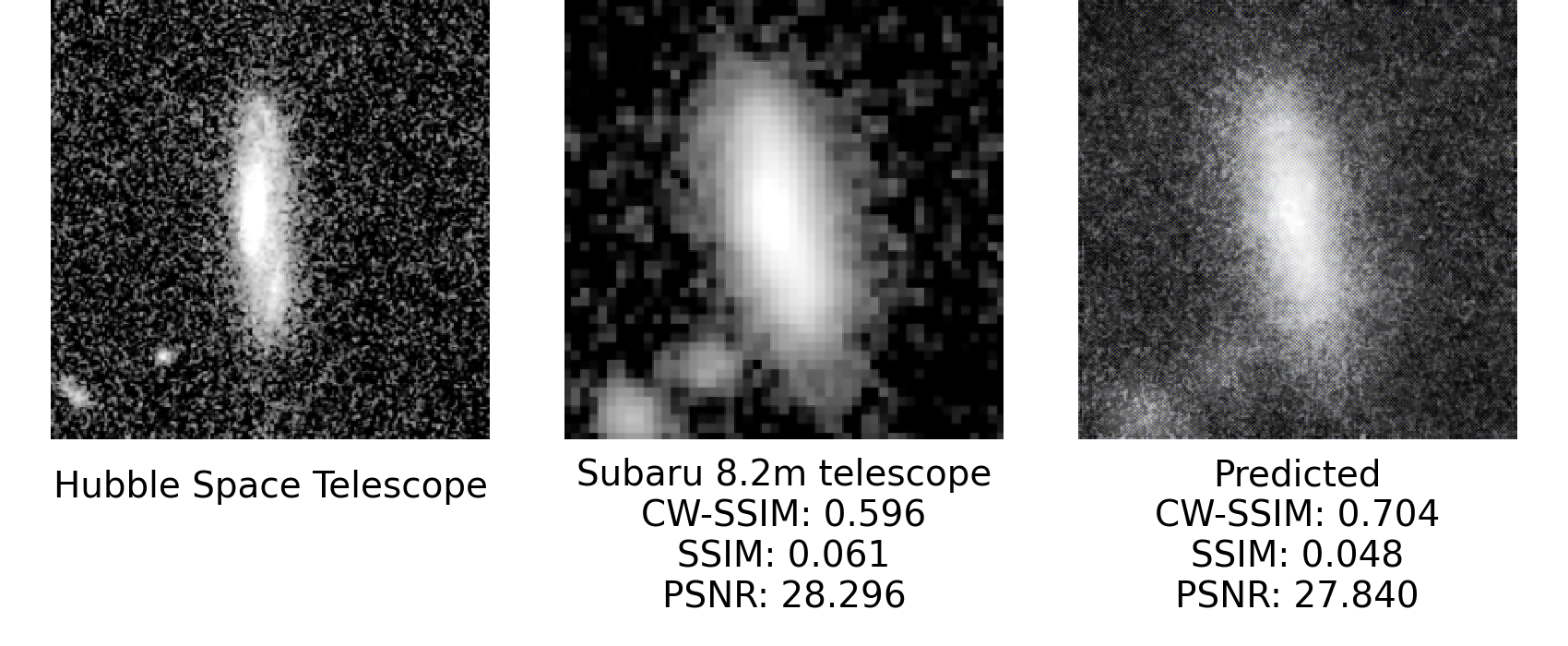}
	(g)
     \end{minipage}
             \vspace{0.025in}
          \hfill{}
      \begin{minipage}[r]{1.0\columnwidth}
         \centering
	\includegraphics[width=\textwidth]{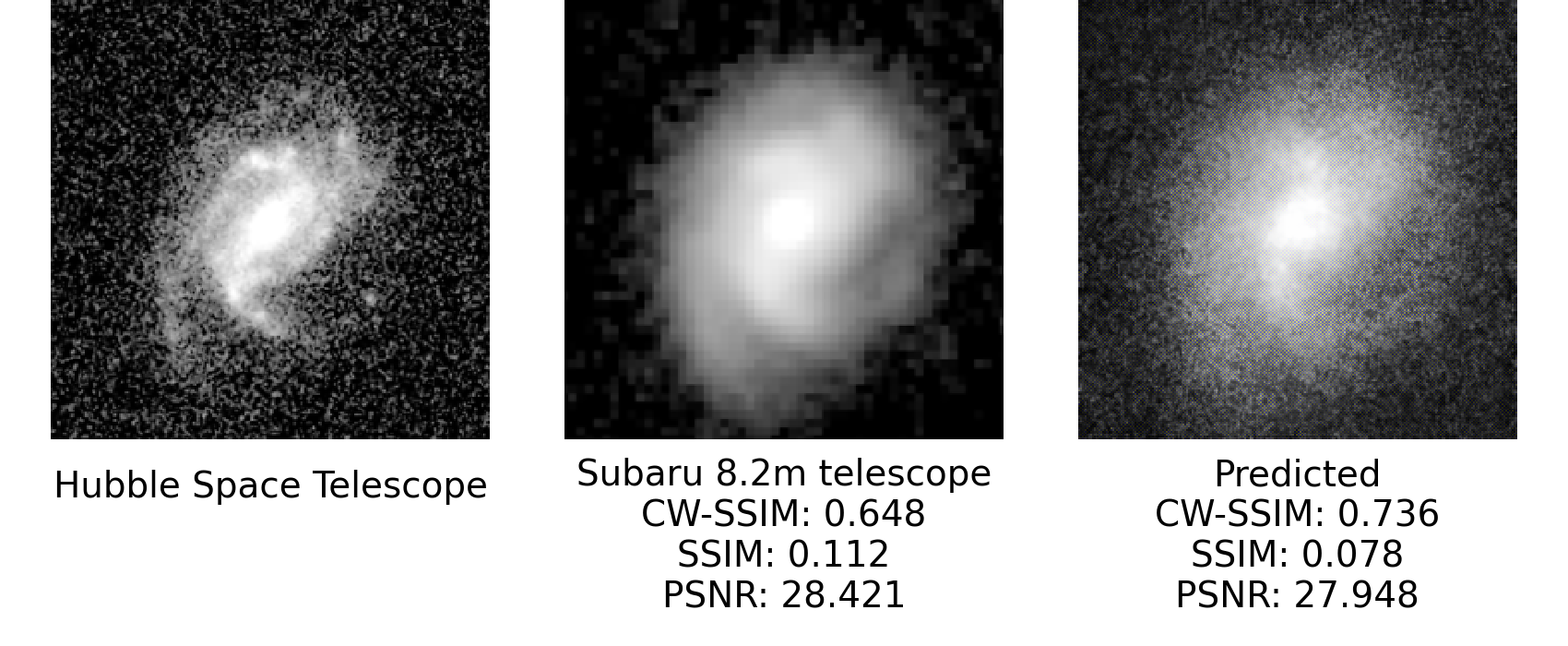}
	(h)
     \end{minipage}
     	        \vspace{0.025in}
	        
\caption{Sample results produced by SeeingGAN. The images are listed in the order of HST, Subaru, SeeingGAN prediction. The SeeingGAN result is obtained by predicting the results from the Subaru image. The CW-SSIM value is obtained by comparing the said image and the HST image, a higher CW-SSIM value indicates that the image is more similar to the HST image.}
\label{section:result}

\end{figure*}

\begin{table*}
\centering
\begin{tabular}{|l|c|c|} 
\hline
Attributes & Subaru - HST & SeeingGAN - HST \\
\hline
PSNR & {28.412 / 28.139 / 28.738} * & 27.913 / 27.820 / 28.237 \\ 
SSIM & {0.084 / 0.017 / 0.261} * & 0.062 / 0.023 / 0.132 \\ 
CW-SSIM & 0.560 / 0.302 / 0.711 & {0.622 / 0.307 / 0.760} * \\ 
\hline

\end{tabular}
\caption{Quantitative results comparison. The results were obtained by averaging PSNR, SSIM, CW-SSIM across the test dataset of 100 images between Subaru-HST and SeeingGAN-HST. The scores are listed in the order of - average result of all the image pairs, minimum result, maximum result. A higher score represents higher similarity. *indicates the better result of the metric between the 2 image pairs.} 
\label{tab:table1}

\begin{tabular}{|l|r|} 
\hline
Attributes & Percentage Improvement of CW-SSIM  \\
\hline
Average	 & 11.15\% \\
Standard Deviation& 3.53\% \\
Minimum	& 1.69\% \\ 
Maximum	& 18.04\% \\ 
\hline
\end{tabular}
\caption{CW-SSIM Metric Improvements. These results were obtained by averaging the CW-SSIM results across the test dataset of 100 images. The CW-SSIM value of SeeingGAN - HST was compared against Subaru - HST.} 
\label{tab:table2}

\end{table*}

\subsubsection*{Complex Wavelet Structural Similarity Index (CW-SSIM)}

We attempted to quantify the model using the above metrics – PSNR and SSIM. However, the predicted model’s PSNR and SSIM metrics were not consistent with the visual improvements we obtained. For most of the images, we can clearly discern more details in the predicted image, but the PSNR and/or SSIM can be much better in the low-resolution image than the predicted image. This behaviour is a known consequence when the compared images that are not geometrically similar because these metrics depends heavily on the pixel location spatially and are sensitive to small scaling, rotation and translation \citep{Sampat2009}. Hence, a variant of SSIM was created – CW-SSIM by \citealp{Sampat2009}. Given two sets of coefficients $c_x = \{c_{x,i}| i = 1,...M\}$ and $c_y = \{c_{y,i}| i = 1,...M\}$ extracted from the same spatial location of the same wavelet sub-band of 2 images being compared,

\begin{equation}
\begin{multlined}
\CWSSIM (c_{x}, c_{y}) = \Bigg( \frac {2\sum _{i=1}^{N}|c_{x,i}||c^*_{y,i}|+K}{\sum _{i=1}^{N}|c_{x,i}|^{2}+\sum _{i=1}^{N}|c_{y,i}|^{2}+K} \Bigg)
 \end{multlined}
\end{equation}

\noindent where $c^*$ is the complex conjugate of $c$, $K$ is a small positive stabilising constant.

The difference between this variant and the vanilla SSIM is that slight geometric distortions in the images will result in more consistent local phase patterns in the complex wavelet transformation domain. In simpler terms, CW-SSIM is more robust against minor geometrical differences and can still deliver SSIM-like numerical results that are perceptually representative. CW-SSIM also ranges alike SSIM, -1 to 1.    

\subsection{Training}
\label{section:training}

We implemented the models using Keras backed via Tensorflow 2.1. The training was done on a single NVIDIA V100 GPU with 600 256x256 images and 100 test images.  We started by training the matched image pairs with our network. Because each layer in the VGG-19 network extracts different abstract features from the images, our initial results showed that our initial selection of loss layer was providing a suboptimal objective function for the model to train on, resulting in images that have weird veils across the images, shown in Fig. \ref{section:loss-layers}. Hence, we iterated different loss layers in the discriminator to obtain a better result, which we ended up with the $conv_{3,1}$ in VGG-19.

Secondly, the model gradually started to add more stars around the galaxy, which we think the model was trying to simulate background stars. However, these stars are not technically correct because they are too small for the blurred image to resolve, so they are probably random noise added by the network to the output images to make them look realistic. Hence, we cropped the images programmatically to fill up the window with the galaxy as much as possible.

\section{Results}
\label{section:results}
\subsection{Individual Cases}
\label{section:qualitative-results}

First, we describe the results for a typical example of predicted images by SeeingGAN in Fig. \ref{section:result-enlarged}. Then we show the results for a number of selected images predicted by SeeingGAN in Fig. \ref{section:result}, because they show a variety of prediction accuracies. Other interesting cases are explained in detail in Appendix \ref{appendix:a}.

As we can see in Fig. \ref{section:result-enlarged}, the predicted image on the right clearly shows the centre and the arms of the spiral galaxy. As compared to the Subaru image, we can only see large blob of stars in the centre, and we cannot easily discern the type of galaxy. The CW-SSIM of the predicted image is 0.612, whereas the original image is 0.580, which shows that the predicted image is more similar to the HST image (due to the 0.032 higher value in CW-SSIM). This means that the image is clearer because it has less seeing effects. The SSIM values are 0.095 and 0.206, respectively, and PSNR values are 27.991 and 28.523 for the predicted image and original image, showing that the predicted image is worse-off according to these metrics. As noted in Section \ref{section:eval_metrics}, both the more common SSIM/PSNR values are highly susceptible to geometrical differences. Hence they were not an excellent candidate to quantify our model in this instance. Furthermore, because our model does not use mean square error as an objective, it is not directly optimised for the PSNR metric. Even though we can qualitatively conclude that the predicted image is clearer, the quantitative metric says otherwise. This behaviour is consistent across most of our dataset. Hence, we can confidently say that the SSIM/PSNR metric is not a good prediction of reconstruction quality for this research. Thus, the CW-SSIM is used throughout this research because it agrees with the qualitative metric more accurately.

By visual inspection, SeeingGAN produces superior results as compared to the original low-resolution image pair. It can be observed that our model enhances the texture of the low-resolution blurred image and introduces finer edges to the images. Fig. \ref{section:result} shows eight examples of deblurred images predicted from our new GAN-based model. These images clearly demonstrate that the fine structures such as clumps and spiral arms can be better seen our new GAN-based model's predicted images compared to the original Subaru-images. For instance, in Fig. \ref{section:result} (a), we manage to show two tail-like structures that may result from the past tidal interaction of other galaxies. In Fig. \ref{section:result} (b), the 2 arms of the spiral galaxy is more apparent as the blurriness between the centre and the arm is removed by SeeingGAN. In image Fig. \ref{section:result} (d) and (h), the predicted image refines the centre of the galaxy and clearly shows the spiral arms of the galaxy as compared to the original low-resolution telescope. 

We can observe that the model increases the CW-SSIM values, which meant that the deblurred image is more similar to the HST-based image. Generally, SeeingGAN is good at deblurring the centre and edges of the galaxies. We can see that the centre of the galaxy is clearer, showing the type of galaxy. As compared to the Subaru image, where the centre of the image is blurred, it is not easy to deduce the type of galaxy.

For some images in Fig. \ref{section:result} and in Appendix \ref{appendix:a}, however, the deblurring effect is not as dramatic as expected. For example, in Fig. \ref{section:result} (c) and (g) \& Appendix \ref{appendix:a} (c) and (e), even though the predicted image has a higher CW-SSIM value, the model did not deblur the images considerably. The model did reduce the pixelation of the source image, but the result is not significant enough to be able to identify the galaxy type. This is, unfortunately a deficiency of this model, and we think this is due to the lack of image data in the source image, i.e., if the image is too blur or too small, there is not enough distinctive features in the source image to help the model predict the output. From our experiments, the model tends to underperform in clumpy like galaxies, presumably due to the lack of significant structural variation. Hence, more work has to be done to identify the limits of this model, and change to the model to improve the output.

\subsection{Statistics of predicted results}

As noted in the section \ref{section:eval_metrics} and Table \ref{tab:table1}, PSNR and SSIM metrics were not able to quantitatively measure the improvements of the predicted output from the model. Looking at the results, if we relied solely on PSNR \& SSIM, the model would produce worse off results. Hence, as discussed in Section \ref{section:eval_metrics}, CW-SSIM was utilised to provide a better measurement of improvement. In Table \ref{tab:table2}, the predicted model can provide an average of 11.15\% better CW-SSIM score over the original Subaru - HST image, and in some instances, a 18\% better CW-SSIM score. As mentioned in section \ref{section:eval_metrics}, a higher CW-SSIM generally equates to a more similar image, and in this case, an increase in similarity to HST is effectively removing the seeing effects of the atmospheric distortion. Additionally, the standard deviation is relatively small in our results, showing the consistency of our model.

To better intuitively explain the physical meaning of the score improvement, in another image deblur paper, DeblurGAN \citep{Kupyn2017}, a 1\% increase in SSIM value increases the object detection F1 score by approximately 4\%.  Although not directly comparable, we can estimate that a 11.15\% increase in CW-SSIM score may increase the object detection F1 score by orders of magnitude.

These results demonstrate that SeeingGAN can convert other existing Subaru images that have no HST counterparts into clearer images. However, it should be stressed that this CW-SSIM performance is only for image pairs from the Subaru and HST image pairs. It is not guaranteed that we can obtain a similar result for other pairs of images, such as VLT-HST or SDSS-HST. Thus, our future study will develop similar a SeeingGAN model for other pairs images from ground and space telescopes.

\section{Discussion}

For the first time, we have demonstrated that many pairs of images from (ground-based) Subaru and (space) HST enable us to develop SeeingGAN that can deblur from Subaru images quite well. This means that astronomers can use SeeingGAN  to (i) deblur Subaru images with no HST counterparts (ii) see the fine structures of galaxies more clearly. Furthermore, this means that one can also develop SeeingGAN using many pairs of images from other ground-based telescopes such as AAT and VLT and from HST. For example,  there are a large number of galaxy images from SDSS \citep{Gunn2006} and GAMA projects \citep{Driver2010}, which can be used to train SeeingGAN if there is a large enough sample of HST counterpart images. It would be reasonable to say that the new architecture developed in the present study (or something similar to the present one) can be used to develop SeeingGAN for other combinations of ground-based telescopes and HST. 

There are several scientific merits of our SeeingGAN in astronomical studies. First, astronomers can see the internal fine structures of galaxies such as spiral arms, tidal tails, and massive clumps more clearly, which can be difficult to see in optical images of distant galaxies from ground-based telescopes. These generated clearer images  by SeeingGAN would assist astronomers to classify galaxies better and discover new internal structures of distant galaxies which otherwise could be difficult to find in original blurred images. For example, it could be possible that distant galaxies classified as S0s with no spirals in original blurred images are indeed spiral galaxies in the debarred images by SeeingGAN. This can influence the redshift evolution of S0 fraction in groups and clusters, discussed in many recent papers (e.g., \citealp{Just2010}). Also,  SeeingGAN can be used for citizen science projects for galaxy classification by the public, e.g., the Galaxy Zoo project. If galaxy images in these projects are blurred (more challenging to classify galaxies),  then the deblurred images generated by SeeingGAN can be easily used for the public galaxy classification instead of the original image. The speed at which SeeingGAN can convert blurred images to deblurred ones is rapid, it is not tricky for SeeingGAN to generate a massive number of deblurred galaxy images.

As shown in Fig. \ref{section:result},  the deblurred images are clearer than the original Subaru images, however, some of them are not dramatically clearer as the HST counterparts. Hence, our future study investigates whether different CNN architectures,  larger numbers of image pairs, and model parameters of SeeingGAN can improve the performance of SeeingGAN. Since the present study has proposed one example of SeeingGAN for a limited number of Subaru-HST image pairs,  it is worth a try for us to investigate different architectures of GAN for a much larger number of image pairs
We plan to use the large number (a million) of Subaru Hyper Suprime-Cam and HST images to test new architectures of SeeingGAN for its better performance. It might be essential for us to use galaxy images from other optical telescopes (e.g., VLT) to confirm that SeeingGAN can be developed from different combinations of ground-based and space telescopes. Although we have focused exclusively on galaxy images in optical wavelengths,  it might be an interesting future study to use galaxy images at other wavelengths from space telescopes (e.g., JWST) to develop new SeeingGAN.

\section{Conclusion}
\label{section:conclusion}

We have demonstrated that SeeingGAN is able to assist astronomers to see fine structures of galaxies such as spiral arms, bars, clumps, tidal tails etc. taken from traditional ground-based telescopes and amplify the details present in the source image to HST-like resolution without the atmospheric distortion with promising quantitative and qualitative results. By only utilising deep learning methods, we can augment and leverage the capabilities of traditional ground-based telescopes without any physical modifications. This proves to be a relatively simple yet effective solution to remove atmospheric distortion from current ground-based telescopes without investing significant resources to install advanced equipment like adaptive optics in the VLT. 

In the current status quo, many Subaru images do not have a clear, HST-like counterpart. Hence the first outcome of this project could be deblurring the massive library of available Subaru images. Since we have successfully developed a GAN-based model to deblur images, we can develop similar models to deblur images from other ground-based telescopes such as SDSS, VLT and Keck, if many pairs of images of these and HST are available. We plant to develop newer SeeingGAN for images from these other ground-based telescopes. Lastly, we would like to iterate that this paper is the first step for deep-learning based image deblurring for space images, and we have further improvements planned to improve and further explore the limits of this family of deblurring methods.

\section*{Data availability Statement}
The data underlying this article will be shared on reasonable request to the corresponding author.

\bibliography{paper}
\bibliographystyle{mnras}

\clearpage
\appendix
\section{\\Additional Results}
\label{appendix:a}

Below are more results generated from SeeingGAN. The images are in the order of HST, Subaru Telescope and the predicted SeeingGAN output. In both the Subaru and SeeingGAN images, we also included the CW-SSIM, SSIM and PSNR values. These values are obtained by comparing the said image to the HST counterpart. A higher score represents a more substantial similarity to the HST image, indicating that the image is clearer. For example, in the first two sets of images below (a) and (b), we can see that the galaxy's centre is revealed, showing a cluster-like galaxy. As compared to the Subaru image,the centre of the image is blurred, and it is not easy categorise the type of galaxy.

\FloatBarrier

\begin{figure}
\centering
\vspace*{-3in}
\includegraphics[width=\textwidth/2-20pt]{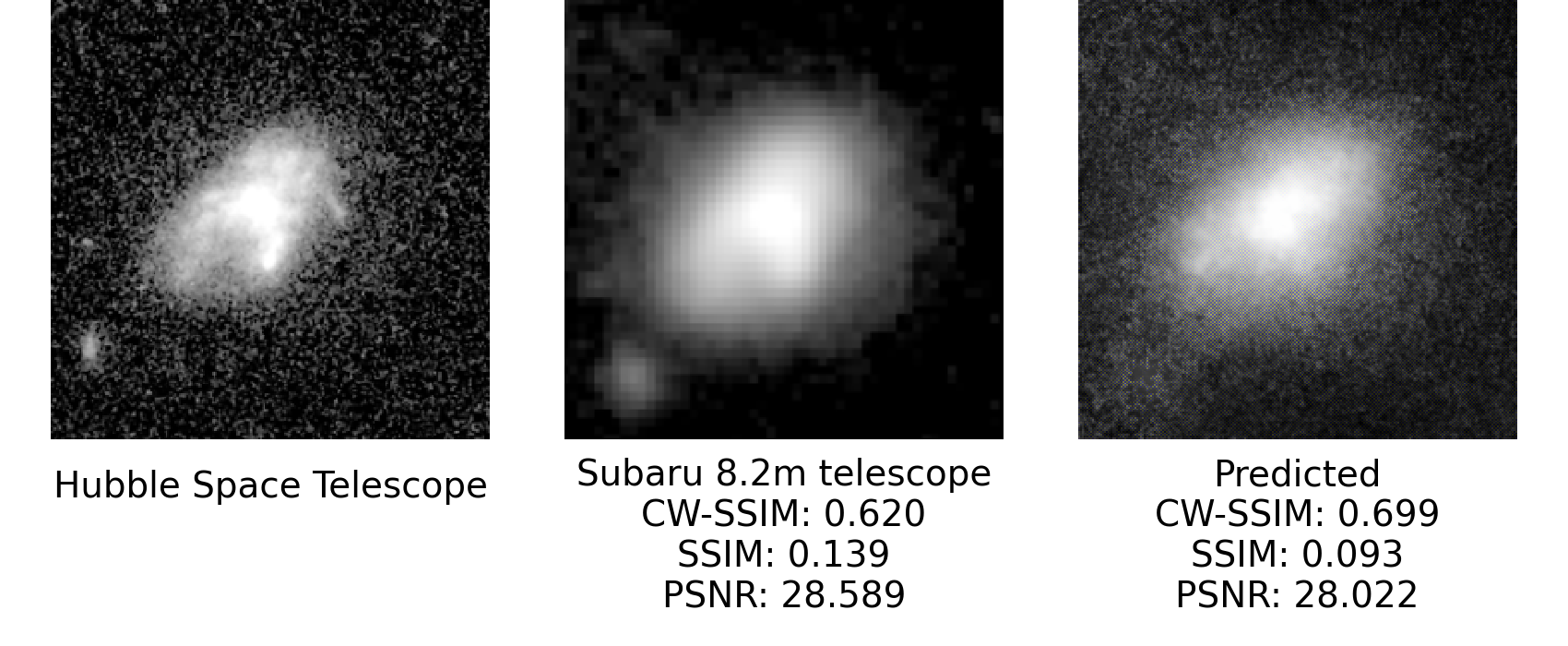}
\vspace*{0.1in}
(a)
\includegraphics[width=\textwidth/2-20pt]{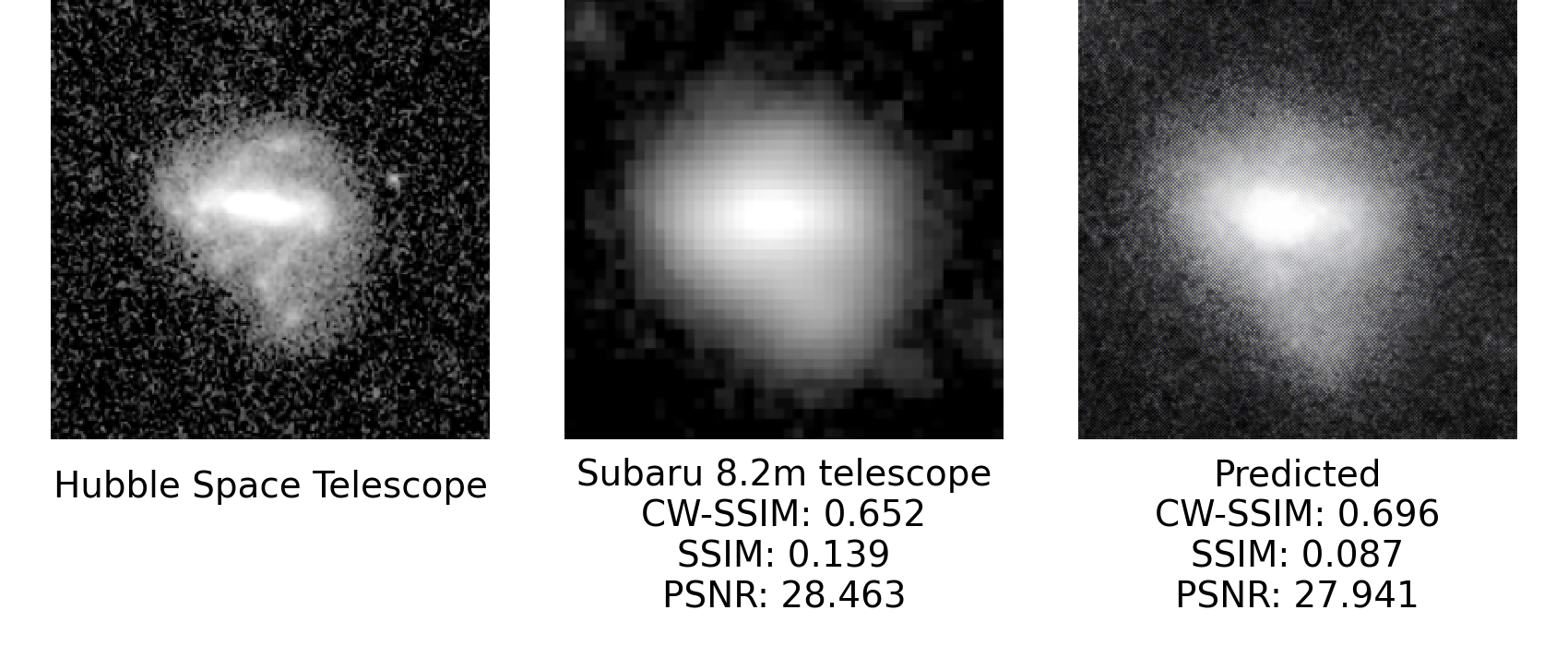}
\vspace*{0.1in}
(b)
\includegraphics[width=\textwidth/2-20pt]{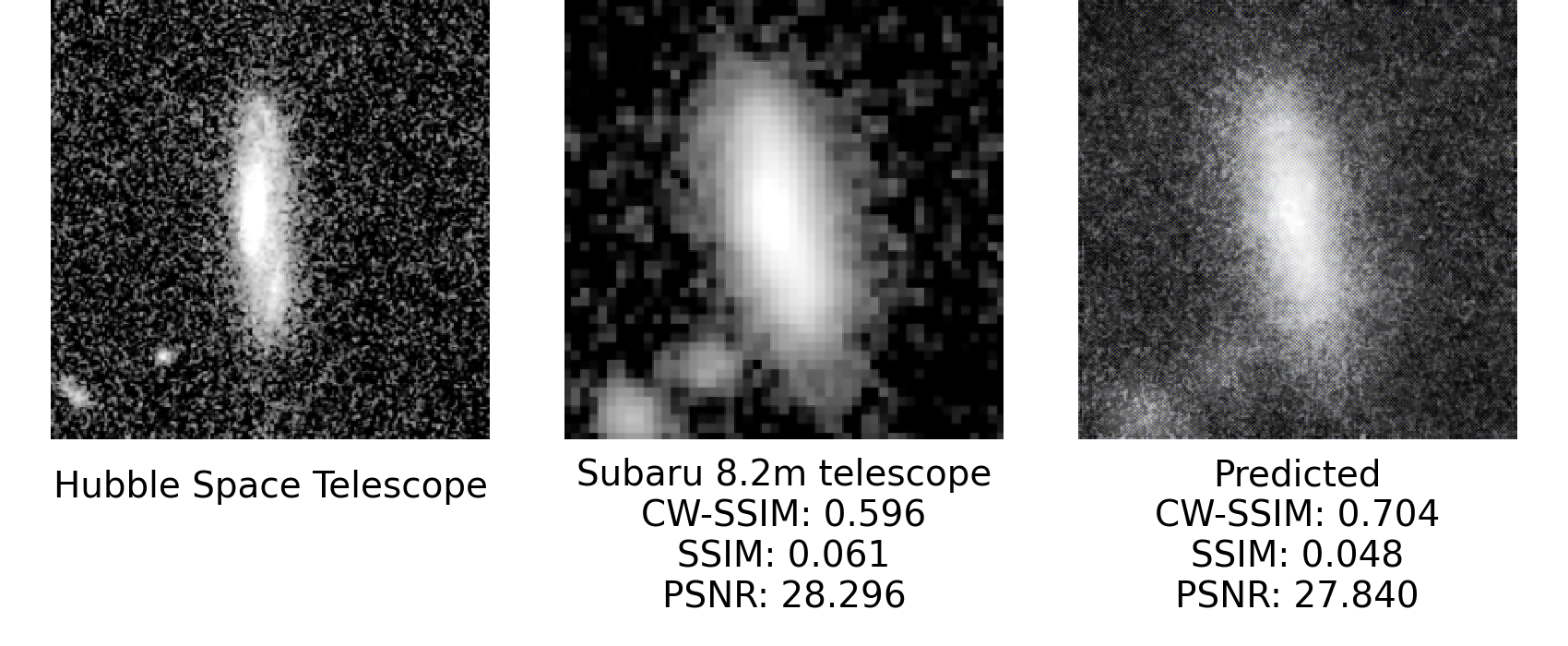}
\vspace*{0.1in}
(c)
\includegraphics[width=\textwidth/2-20pt]{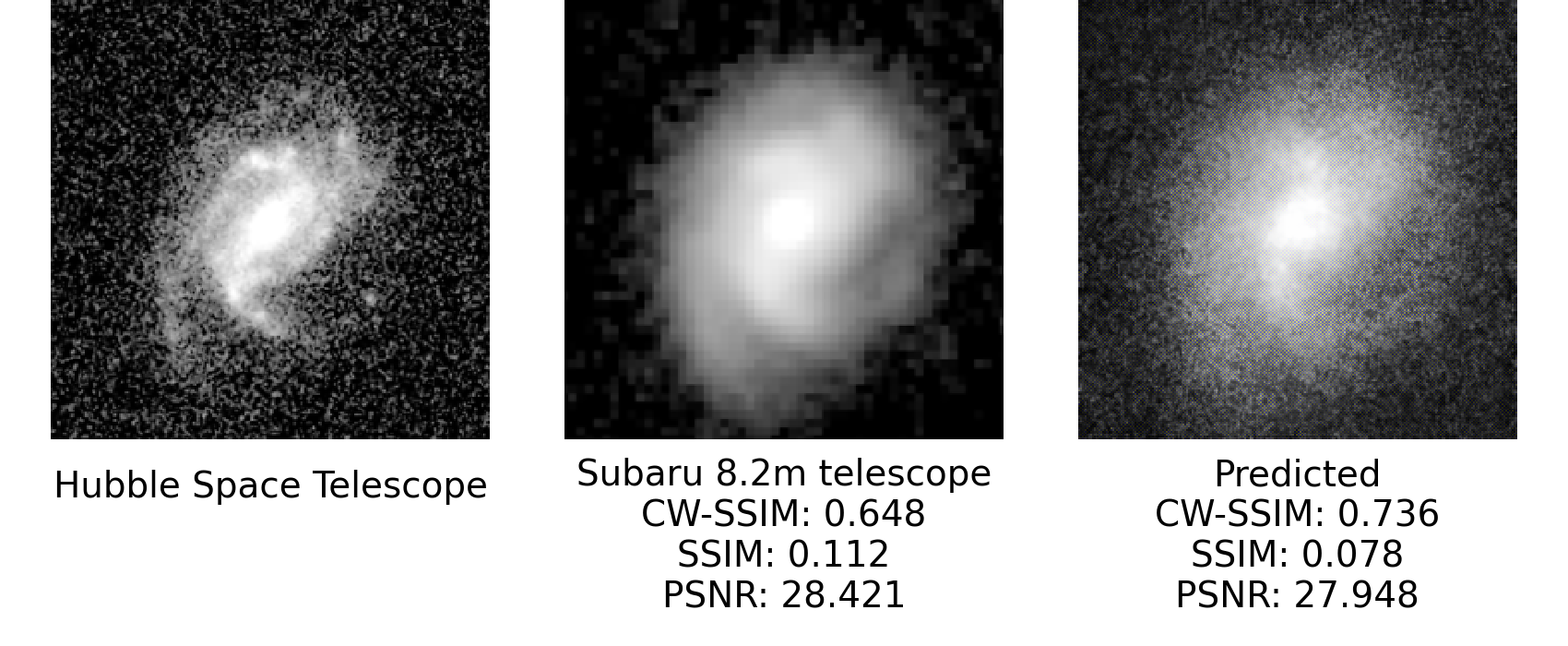}
\vspace*{0.1in}
(d)

\vspace*{0.5in}
\end{figure}

\begin{figure}
\centering
\includegraphics[width=\textwidth/2-20pt]{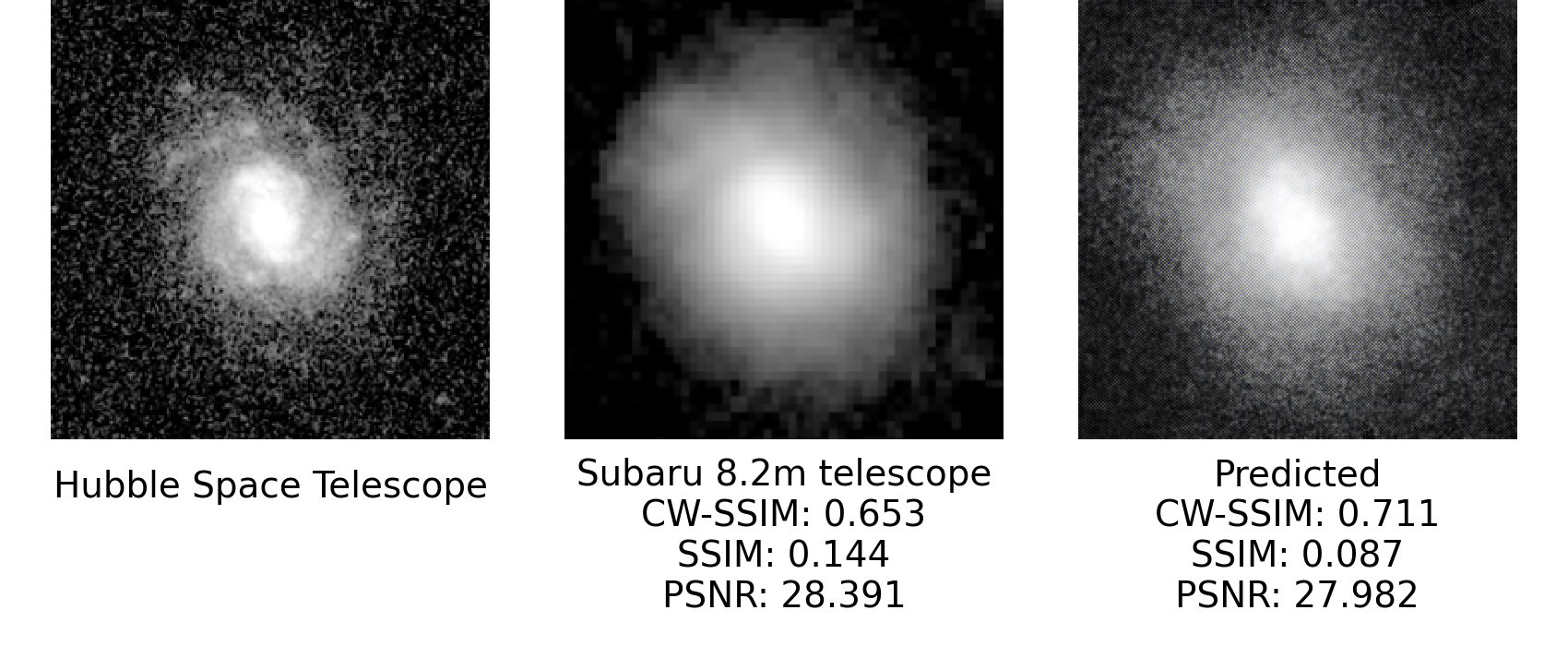}
\vspace*{0.1in}
(e)
\includegraphics[width=\textwidth/2-20pt]{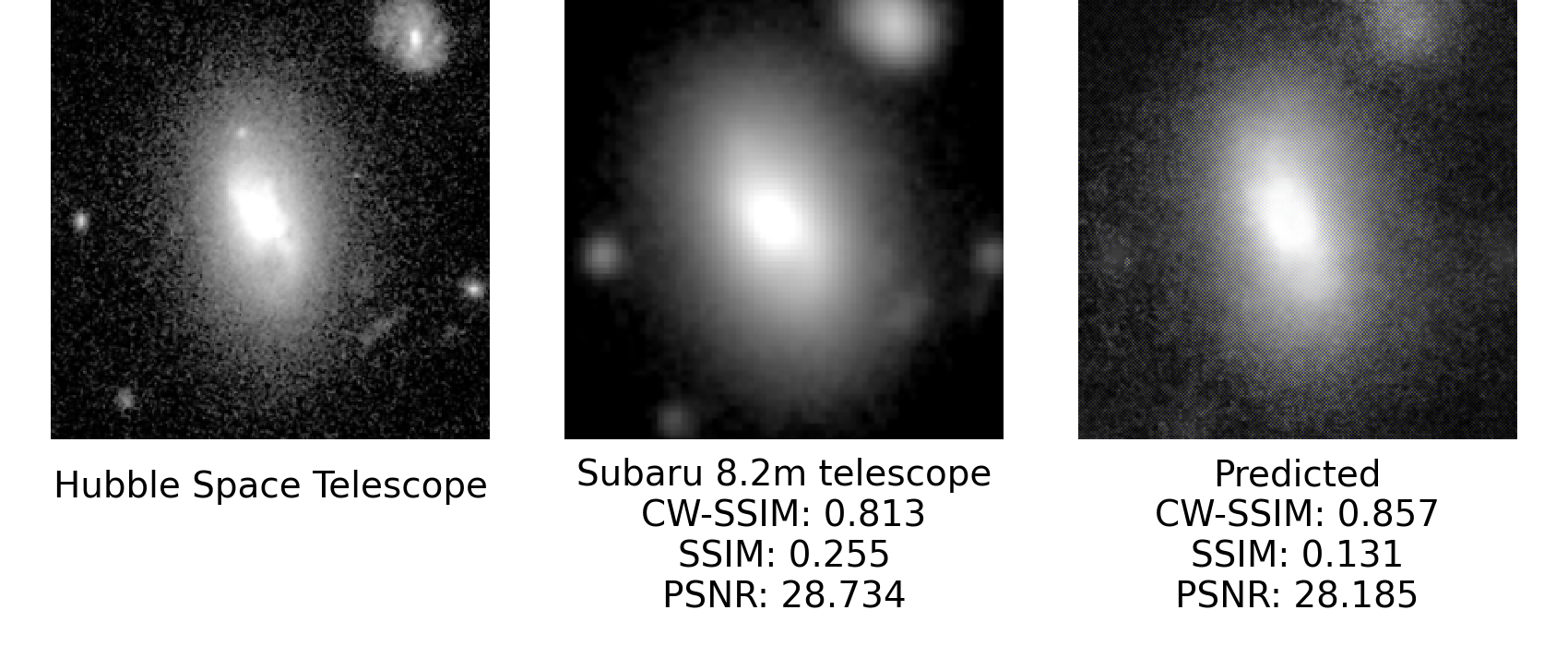}
\vspace*{0.1in}
(f)
\end{figure}

\end{document}